\documentclass[prd,floatfix]{revtex4}

\def\be{\begin{equation}}
\def\ee{\end{equation}}
\def\la{\label}
\def\bea{\begin{eqnarray}}
\def\eea{\end{eqnarray}}
\def\non{\nonumber}
\def\ci{\cite}
\def\la{\label}
\def\bib{\bibitem}

\def\lm{\lambda}
\def\lu{\lambda_1}
\def\ld{\lambda_2}
\def\lt{\lambda_3}

\def\le{\left}
\def\ri{\right}

\def\gm{\gamma}

\def\vp{\varphi}
\def\Omp{\Omega_\phi}

\def\Om{\Omega}

\def\rp{\rho_\phi}

\def\wp{w_\phi}

\def\weff{w_{eff}}
\def\s8{\sigma_8}

\def\fr{\frac}
\def\pp{\partial}
\def\pu{\pp_\mu}
\def\pU{\pp^\mu}

\def\non{\nonumber}

\def\Omp{\Omega_\phi}

\def\r{\rho}

\def\wvp{w_\vp}
\def\Omvp{\Omega_\vp}
\def\rp{\rho_\phi}

\def\rb{\rho_b}
\def\wb{w_b}

\def\Ob{\Omega_b}

\def\wpe{w_{\phi eff}}
\def\wvpe{w_{\vp eff}}

\def\gb{\gamma_b}

\def\wp{w_\phi}

\def\we{w_{eff}}

\def\sdt{\sqrt{1-\dot T^2}}

\def\pt{p_T}
\def\rt{\rho_T}
\def\wt{w_T}
\def\Omt{\Omega_T}
\def\wte{w_{T eff}}
\def\weu{w_{1 eff}}
\def\wed{w_{2 eff}}
\def\gmed{\gamma_{2\; eff }}
\def\gmeu{\gamma_{1\; eff }}
\def\gme{\gamma_{eff}}

\usepackage{graphicx}

\begin{document}

\title{Interacting Tachyon: generic cosmological evolution for  a tachyon and a  scalar field}

\author{A. de la Macorra and U. Filobello}
\affiliation{Instituto de F\'{\i}sica,\\
Universidad Nacional Aut\'onoma de M\'exico,
\\ Apdo. Postal 20-364, 01000 D.F.  M\'exico }

\begin{abstract}

We study the cosmological evolution of a tachyon  scalar field $T$
with a Dirac-Born-Infeld type lagrangian and potential $V(T)$
coupled to a canonically normalized  scalar field $\phi$ with an
arbitrary interaction term $B(T,\phi)$ in  the presence of a
barotropic fluid $\rb$, which can be matter or radiation. The force
between the barotropic fluid and the scalar fields is only
gravitational. We show that the dynamics is completely determine by
only three parameters $\lu = -  V_T/ V^{3/2}, \ld = -  B_T /
B^{3/2}$ and $\lt =-B_{\phi} / B$. We determine  analytically  the
conditions for $\lm_i$ under which the energy density of $T$,
$\phi$ and $\rb$ have the same redshift. We study the behavior of
$T$ and $\phi$ in the asymptotic limits for $\lm$ and we show the
numerical solution for different interesting cases.

The effective equation of state for the tachyon field changes due
to the interaction with the scalar field and we show   that it is
possible for a tachyon field to  redshift as matter in the absence
of an interaction term $B$   and   as radiation when $B$ is turned
on. This result solves then the tachyonic matter problem.

\end{abstract}


\maketitle

\section{Introduction}

String tachyon fields $T$ are the lowest energy state
in  unstable Dp-brane or  brane-antibrane systems \ci{DB2}.
Since they represent the
low energy limit of string-brane models the
phenomenology of the tachyon field is
important and
a great amount of work has been invested in studying
the dynamics of tachyon field \ci{DB2},\ci{slowroll}.
In the case of a Dp-brane systems in  string theory,  the potential
$V(T)$ has been conjectured  to be tachyonic at the
origin $T=0$ \ci{DB2},  i.e. the potential has a maximum at the origin
with a negative  mass square, $m^2=V_{TT}<0 $.

The tachyon field has a Dirac-Born-Infeld type lagrangian and
therefore it does not have canonical kinetic terms. This implies
that the naive prescription for the mass of tachyon as the
$m^2=V_{TT}$ does not hold in this case  \ci{tach.mio} nor can we assume that
the evolution of $T$ is to reach the minimum of the potential
$V(T)$, as for a standard scalar field, since $V(T)$ does
not correspond to the true    potential for $T$.

It is well known that the tachyon field has
an equation of state parameter $-1\leq w \leq  0$ \ci{slowroll}. For
string motivated potentials, e.g. $V\simeq e^{-T^2/2}$,
the late time  behavior gives   $w=0$ and  $T$ behaves
as matter. This is the "matter problem" because  $T$
can easily dominate the universe well before radiation-matter equality
since it redshifts slower than radiation.
So, if $T$ is present at early times it should necessarily
decay into other particles to avoid the matter problem.

In this letter we study the generic cosmological evolution
of a scalar field $T$ with a Dirac-Born-Infeld type lagrangian
coupled to a canonical scalar field through
an arbitrary interaction term $B(T,\phi)$ in the presence a barotropic
fluid, which can be matter or radiation.
We will call the  field $T$ as the tachyon field
even though we do not constrain ourselves to a potential $V(T)$
with $m^2<0$ at the origin, so our results are valid
for any potential $V(T)$.
We show that all models dependence is given in terms of three parameters
$\lu = -  V_T/ V^{3/2},
\ld = -  B_T / B^{3/2}$ and
$\lt =-B_{\phi} / B$. We determine the dynamical
equations and obtain the attractor solutions as a function
of these $\lm_i$ parameters. We show in this {\it letter} that it is  possible
for a tachyon field to
redshift as matter in the absence of an interaction
term $B$ and to  redshift at late times as radiation due to
the interaction term. Therefore, the interaction term
solves the tachyonic matter problem.

This {\it letter} is organized as follows. In section \ref{sf}
we set up the framework for the cosmological evolution
of two scalar fields, a tachyon and a canonical
scalar field, with an arbitrary potential in the presence
of a barotropic fluid.
In section \ref{gda} we derive the dynamical first order
differential  equations and we show that the system
is determined by only three parameters. In section \ref{cs}
we calculate the critical points.   In section \ref{lim}
we study different asymptotic limits and we present
a discussion on specific particle physics motivated examples in section
\ref{ppm}. Finally in section \ref{ex} we give some interesting
examples and
 we present our conclusions in section \ref{con}.

\section{Coupled Tachyon and Scalar Field}\la{sf}

Our starting point is a universe  filled with two
scalar fields $T,\phi $ and a barotropic  energy density $\rb$,
which can be either matter $\wb=0$ or radiation $\wb=1/3$.
We will assume that the  scalar fields interact
via a potential $B(T,\phi)$ while there is only gravitational
interaction between these fields and the barotropic fluid. This work generalizes
that
of  a single scalar field and a barotropic fluid \ci{mio.gen},
a tachyon field and a barotropic fluid\ci{tach.tr} and two single scalar field
with arbitrary potential and a barotropic fluid \ci{mio.2gen}.

One of this
scalar fields, namely $T$, is a tachyon field which is motivated
by string theory and D-branes, while the other scalar field $\phi$ is
a standard canonical field.
We take the following Lagrangian for the scalar fields $\phi$ and
$T$ \ci{DB2}
\be\la{LT}
L=- V(T)\sqrt{1-\pu T\pU T}+ \fr{1}{2}\pu \phi\pU\phi   - B(\phi,T)
\ee
where the tachyon is given by a Dirac-Born-Infeld type lagrangian with a
potential $V(T)$ and $B(\phi,T)$ corresponds to the
interaction between $\phi$ and  the tachyon $T$. The
potentials  $V(T)$ and $B(\phi,T)$ are completely arbitrary.
The equation of motion of $\phi$ and
$T$ for a spatially flat Friedman--Robertson--Walker  (FRW)
universe are
\bea\la{dp}
\ddot\phi+3H\dot\phi + B_\phi&=&0\\
\fr{\ddot T}{1-\dot T^2}+3H\dot T +\fr{V_T}{V}+\fr{B_T}{V}\sdt&=& 0
\la{dvp}\eea
where  the subindex in $V$ and $B$ is defined as $V_T\equiv \pp V/\pp T,
 B_\phi\equiv\pp B/\pp\phi$ and
$B_T\equiv\pp B/\pp T$.
The Hubble parameter $H\equiv\dot a/a$ is
\be\la{H}
3H^2=\rho =\rp+\rt+\rb
\ee
where we have taken $8\pi G\equiv 1$ and $\rho$ is the total energy density,
$\rb$   the barotropic fluid and $\rp,\rt$ are defined as
\be\la{rp}
\rp \equiv\fr{1}{2}\dot\phi^2+  V(\phi),\hspace{.5cm} p_\phi \equiv\fr{1}{2}\dot\phi^2 -
V(\phi)
\ee
for $\phi$ and
\be\la{rt}
\rt \equiv\fr{V}{\sdt} , \hspace{.5cm} \pt\equiv V\sdt.
\ee
for  $T$ with $p_\phi, p_T$  the pressure of $\phi, T$, respectively.
We define the ratio of energy densities as $\Omt\equiv \rt/3H^2$, $\Omp\equiv \rp/3H^2$.
Using eqs.(\ref{rp}) and (\ref{rt}) we can rewrite the dynamical
eqs.(\ref{dp}) and  (\ref{dvp}) in terms of the energy densities as
\bea\la{dr}
\dot\rp+3H\rp(1+\wp) &=&B_T\,\dot T =\delta   \non\\
\dot\rt+3H\rt(1+\wt) &=& -B_T\,\dot T=-\delta   \\
\dot\rb+3H\rb(1+w_b) &=&0\non
\eea
where we have  included the evolution of the barotropic fluid $\rb$ and
\be\la{d}
\delta\equiv B_T\,\dot T
\ee
defines the interaction term.
The equation of state parameters are given by
 \be\la{wt}
\wp\equiv\fr{p_\phi}{\rp}= \fr{\fr{1}{2}\dot\phi^2-V}{\fr{1}{2}\dot\phi^2+V},\hspace{1cm}
 \wt\equiv\fr{p_T }{\rt}= -1+\dot T^2.
\ee
In order to have a real energy density for the tachyon
we require $0\leq \dot T^2\leq 1$ and from eq.(\ref{wt}) we
see  that the equation of state parameter for $T$ is constraint
to $-1\leq \wt\leq 0$.
The time derivative of $H$ is given by
\be\la{dH}
\dot H =-\fr{1}{2}\le(\rp+\rt+\rb+p_\phi+p_T+p_b \ri)=-\fr{1}{2}\le(\dot\phi^2+\rt \dot T^2 +\rb(1+w_b)  \ri).
\ee

\subsection{Effective   Equation of State}\la{eff}

From  the last two terms in eq.(\ref{dvp}) we can define a
 $T$-derivative of an  effective potential $V_{eff}$ as
 \be\la{vv}
 \fr{dV_{eff}}{dT}\equiv (1-\dot T^2)\le(\fr{V_T}{V}+\fr{B_T}{V}\sdt\ri)
 =\fr{V(V_T +B_T \sdt}{\rt^2}
 \ee
 where we have used $\rt=V/\sdt$. The dynamics of the tachyon field gives
a  vanishing  $dV_{eff}/dT$,  i.e. $dV_{eff}/dT=(V_T +B_T \sdt\;)V/\rt^2=0$.
 It is important to
 include in eq.(\ref{vv}) the multiplicative factor $(1-\dot T^2)$.
 The dynamics of a scalar field is to minimize the potential (i.e.
 the derivative of the potential w.r.t. the scalar field should
 vanish) and in the absence of the interaction term $B$ the dynamics
 of the tachyon field does not give a vanishing $V_T/V$
 but it has a vanishing $(1-\dot T^2)V_T/V$. For
 example for a typical tachyon potential   $V\sim e^{-T^2/2}$
one has   $V_T/V=-T$ and the solution to eq.(\ref{dvp}) gives
 $\dot T^2=1, T\rightarrow\infty $ with $V_T/V \rightarrow -\infty$
 but $(1-\dot T^2)V_T/V=V_T V/\rt^2=0$. This shows that
 $V_T/V$ is not the $T$-derivative of the  potential for the tachyon $T$.

At vanishing $dV_{eff}/dT$, eq.(\ref{dvp}) becomes $\ddot T +3H\dot T (1-\dot T^2)=-dV_{eff}/dT=0$
and it has   two solutions:
\be\la{s1}
 V=0, \hspace{1cm}\ddot T =  0, \hspace{1cm}  \dot T^2=1
\ee
and
\be\la{sdt}
 V_T =-B_T\sdt,\hspace{1cm} \fr{\dot T}{\sdt} =  k\; a(t)^{-3}
\ee
with $k$ an integration constant. The solution in eq.(\ref{s1})
necessarily implies $V=0$ since $\Omt=V/(3H^2\sdt)$ must be smaller than
one while eq.(\ref{sdt}) gives at late times $\dot T=0$ and $V_T=-B_T$.

In order to define an effective equation of state it is
convenient to  rewrite eqs.(\ref{dr}) as
\be\la{rw}
\dot\rp =
-3H\rp(1+\wpe ), \hspace{1cm} \dot\rt = -3H\rt(1+\wte)
\ee
with
the effective equation of state defined by
\be\la{weff}
\wpe
\equiv  \wp - \fr{B_T\dot T}{3H\rp}, \hspace{.5cm} \wte \equiv
w_b+ \fr{B_T\dot T}{3H\rt}.
\ee
We see from eqs.(\ref{rw}) that
$\wpe,\wte$ give the complete evolution of $\rp$ and $\rt$. For
$B_T\dot T>0$ we have $w_{ eff}<\wp$ and the fluid $\rp$ will
dilute slower than without the interaction term  (i.e. $B_T\dot
T=0$) while $\rt$ will dilute faster since $\wte<\wvp$. Which
fluid dominates at late time will depend on which effective
equation of state is smaller. The difference in eqs.(\ref{weff})
is \ci{mio.gen}
\be\la{dw} \Delta w_{eff}\equiv \wte-\wpe=\Delta w
- \Upsilon
\ee
with $\Delta w\equiv \wt-\wp$ and  $\Upsilon$
defined as
\be\la{Us}
\Upsilon=  \fr{\delta}{3H} \le(
\fr{\rp+\rt}{\rp\rt}\ri)=  \fr{B_T\dot T}{3H} \le(
\fr{\rp+\rt}{\rp\rt}\ri)=  \fr{B_T\dot T}{9H^3} \le(
\fr{\Omp+\Omt}{\Omp\;\Omt}\ri)
\ee
while the sum gives
\be\la{sw}
\Omt
\wte+\Omp\wpe= \Omt \wt+\Omp\wp.
\ee Clearly the relevant quantity
to determine the relative growth is given by $\Upsilon$ and if
$\Upsilon < \Delta w$
 we have  $\Delta w_{eff}>0$ (i.e. $\wte>\wpe$) and $\rp$ will dominate the universe at late times
 with $\Omp=1, \Omt=0$.
 For $\Upsilon > \Delta w$ we have  $\Delta w_{eff}<0$ and  $\wte<\wpe$
 with $\rt$ prevailing and   $\Omt=1, \Omp=0$.
If both fluids have the same redshift, i.e.
$\Delta w_{eff}= 0$ and $\weff\equiv\wte=\wpe$
eq.(\ref{sw}) gives
\be\la{OO}
\fr{\Omt}{\Omp}=\fr{\wp-\weff}{\weff-\wt}
\ee
and for $\wp<\wt$ we have
\be
\wp\leq\; \wpe=\wte= \fr{\wt\Omt+\wp\Omp}{\Omt+\Omp} \;\leq \wt,
\ee
while for $\wp>\wt$ we get
\be
\wp\geq\; \wpe=\wte= \fr{\wt\Omt+\wp\Omp}{\Omt+\Omp} \;\geq \wt,
\ee
i.e. the effective equation of state is constraint between $\wp$ and $\wt$.
In the limit of no interaction $\delta=B_{\phi}\dot\phi=0$
 we get  $\Upsilon=0$ and
$\Delta \weff=\Delta w$ and depending in which term is smaller $\wp$ or $\wt$
we will have either  $\rp$ or $\rt$ dominating the universe  at late times.

\subsection{Canonical Scalar Field}\la{csf}

In the limit of having only a canonical scalar
field $\phi$ and a barotriopic fluid, the  potential $B$ depends only
on $\phi$.   If the vacuum expectation value of $\phi_{min}$ has a  finite
 value, i.e.
the potential is of the type $B=b_o\phi^n$,
 then  the scalar field $\phi$  oscillates around its
vacuum expectation value (v.e.v.).
If the scalar field has a non
zero  mass  or if the potential $B$  admits a Taylor expansion
around $\phi_{min}$ then, using the H$\hat { o}$pital rule, one
has  lim$_{t \rightarrow \infty} |B_\phi/B| =\infty$ and
 the energy density $\rp$ redshifts with
$\wp=(n-2)/(n+2)$, i.e. $\wp=0,1/3$ for $n=2,4$ \ci{mio.gen}.
On the other hand, if $\phi_{min}=\infty$ then $\phi$
will not oscillate and  $|B_\phi/B|$ will approach either zero,  a finite
constant  or  infinity. Only in the case $|B_\phi/B|$
going to zero or a constant smaller than $\sqrt{2}$ will
the universe accelerate at late times \ci{mio.gen}.

\subsection{Tachyon Scalar Field}\la{tsf}

If we have only a tachyon field in the presence of a barotropic
fluid, the late time attractor solution were studied in \ci{tach.tr}.
A tracking solution
with constant $\lu\equiv - V_T/V^{3/2}$ is given for
a potential $V=V_o/T^2$.
In this case one finds $\dot T = y_1\lu/\sqrt{3}$ and $y_1^2=V/3H^2=(\sqrt{\lu^4+36}-\lu^2)/6$
with $w_T=-1+\dot T^2=-1+(\sqrt{\lu^4+36}-\lu^2)\lu^2/18$ if $w_b>w_T$
and $ w_T=w_b$ if $w_b<-1+(\sqrt{\lu^4+36}-\lu^2)\lu^2/18$ with
$(x_1,y_1)=(\gb,\pm\sqrt{3\gb}/\lu)$.
\ci{tach.tr}. Clearly a  $|\lu|\ll 1$ gives an equation of
state $w_T\approx -1$ and an accelerating universe.

\section{Generic Dynamical Analysis}\la{gda}

To determine the attractor solutions of the differential equations
given in eqs.(\ref{dp}) and (\ref{dvp}) or (\ref{dr}) it is useful to make the following
change of variables
\bea\la{xy}
x_1 &\equiv&  \dot T,\hspace{1.5cm} y_1 \equiv  \fr{ 1}{H}\sqrt{ \fr{V(T) }{   3} }\\
x_2 &\equiv & \fr{\dot \phi }{ \sqrt{ 6} H},\hspace{1cm} y_2 \equiv
\fr{1}{H} \sqrt{ \fr{B(T,\phi) }{  3} }
\eea
and eqs.(\ref{dr})
and (\ref{dH})  become a set of dynamical differential  equations of first order
\bea \la{cosmo1}
x_{1N}&=& -\le(1-x_1^2\ri)\le(3x_1-\sqrt{3}\,\lu y_1-\sqrt{3(1-x_1^2)}\;\ld\; \fr{y_2^3}{y_1^2} \ri)\non  \\
y_{1N}&=& -\fr{H_N}{H}\; y_1-\fr{\sqrt{3}}{2}\;\lu \;x_1\; y_1^2 \\
x_{2N}&=& -\le(3+\fr{H_N}{H}\ri) x_2+\sqrt{3 \over 2}\;\lt \;y_2^2  \non      \\
y_{2N}&=& -\fr{H_N}{H}\;y_2-\fr{\sqrt{3}}{2}\;\ld\; x_1 \;y_2^2 -\sqrt{\fr{3}{2}}\;\lt\; x_2\; y_2 \non \\
\fr{H_N}{H}&=& -{3 \over 2} \le(  \Om_1\gm_1+\Om_2\gm_2 + \Ob\gb    \ri)=-{3 \over 2} \le( x_1^2\Om_1+ 2x_2^2+ \Ob\gb    \ri)
\la{hh}\eea
where
$N$ is the logarithm of the scale factor $a$, $N \equiv ln (a)$, $ \gb \equiv 1+w_b$,
$\gm_1\equiv 1+w_1$, $\gm_2\equiv 1+w_2$ and $f_N\equiv df/dN$ for $f=H,x_i,y_i\;(i=1,2)$, $\Ob=1-\Om_1-\Om_2$
 and
\be\la{lm}
\lu(N) \equiv - \fr{V_T }{ V^{3/2}},\;\;\;\ld (N) \equiv - \fr{B_T }{ B^{3/2}},\;\;\;\lt (N) \equiv - \fr{B_{\phi} }{ B}.
\ee
Notice that all model dependence in eqs.(\ref{cosmo1}) is through
the three quantities $\lambda_i (N), i=1,2,3$ and the constant parameter $\gb=1+\wb$.
The last equation of (\ref{cosmo1}) is constraint between $-3 \leq H_N/H \leq 0$ for all values of $x_i,y_i$ and $\gb$,
it takes the value $-3$ when the universe is dominated by the kinetic energy
$ x_2^2=1$ and therefore $\Ob=\Om_1=0$  while it becomes $H_N/H=0$ when $x_2=x_1=\Ob=0$ and
the universe is dominate by a
constant potential $ y_1^2+y_2^2=1$.
The set of equations given in eqs.(\ref{cosmo1}) give the evolution of two scalar
fields $\phi,T$, a tachyon and a canonical scalar field, with arbitrary potentials
in the presence of a barotropic (perfect) fluid with equation of state $w_b=1-\gb$.
If we do not want to
consider the contribution from the  barotropic fluid we can easily
 take the limit
$\gb=0$ in eqs.(\ref{hh})   since all contribution form $\rb$ is given in $H_N/H$ via
the term $\Ob\gb$. For $\Ob\neq 0$ we will assume a barotropic fluid with $0<\gb<2$ and $\gb=1$ for
matter while $\gb=4/3$ for radiation.

In terms of $x_i,y_i$ we have
\bea
\Omt=\Om_1=\fr{\rt}{3H^2}&=&  \fr{y_1^2}{\sqrt{1-x_1^2}} ,
\hspace{1cm} \fr{p_T}{3H^2}= - y_1^2 \sqrt{1-x_1^2} \\
\Omp=\Om_2=\fr{\rho_\phi}{3H^2}&=&x_2^2+y_2^2,\hspace{1.2cm}
\fr{p_\phi}{3H^2}= x_2^2-y_2^2
\eea
and
\be
w_1\equiv \wt = \fr{\pt}{\rt}=  -1+x_1^2,\hspace{1cm} w_2\equiv
\wp= \fr{p_\phi}{\rp} =\fr{x_2^2-y_2^2}{x_2^2+y_2^2}
\ee
with  $\gm_1=1+w_1=x_1^2$, $\gm_2=1+w_2=2x_2^2/\Om_2$.
The interaction term  defined in eq.(\ref{d}) and (\ref{Us}) are now
\bea\la{B1}
\delta &=& B_T \dot T= - 3^{3/2} H^3\, \ld   x_1 y_2^3\\
\Upsilon &=&-  \fr{ \ld   x_1 y_2^3}{\sqrt{3} } \le(
\fr{\Om_1+\Om_2}{\Om_1 \;\Om_2}\ri)
\eea
giving an  effective equation of state parameters
defined  in eqs.(\ref{weff}) as
\bea\la{weff2}
\weu\equiv \wte  &=&   w_1 -\;  \fr{\ld\,x_1y_2^3}{\sqrt{ 3}\Om_1}=
\fr{\sqrt{ 3}\,\Om_1 w_1- \ld\,x_1y_2^3}{\sqrt{ 3}\Om_1}\\
\wed\equiv\wpe &=&  w_2+\;   \fr{\ld \, x_1y_2^3}{\sqrt{ 3}\Om_2}
=\fr{\sqrt{ 3}\Om_2 w_2 +   \ld \, x_1\,y_2^3}{\sqrt{ 3}\Om_2}
\eea
and
\be\la{gme}
\gmeu\equiv 1+ \weu,\hspace{1cm} \gmed \equiv 1+ \wed, \hspace{1cm} \gm_{b eff} \equiv \gb.
\ee
The acceleration
of the universe is given by
\be\la{ac}
\fr{\ddot a}{a} = H^2(1+ \fr{H_N}{H}) = -\fr{H^2}{2}\le(\Ob(1+3\wb)+\Om_1(3x_1^2-2)+4 x_2^2-2y_2^2  \ri )
\ee
where we have used $\dot H=H H_N$ and eq.(\ref{hh}).
Clearly acceleration will occur if the universe is dominated by
the potential $ y_2^2 =B/3H^2 $ or $\Om_1$ with $x_1^2<2/3$, i.e.
 for   $y_2^2+\Om_1(1-3x_1^2/2)> \Ob(1+3\wb)+4x_2^2$.

\begin{figure}[htp!]
\begin{center}
\includegraphics[width=6.5cm]{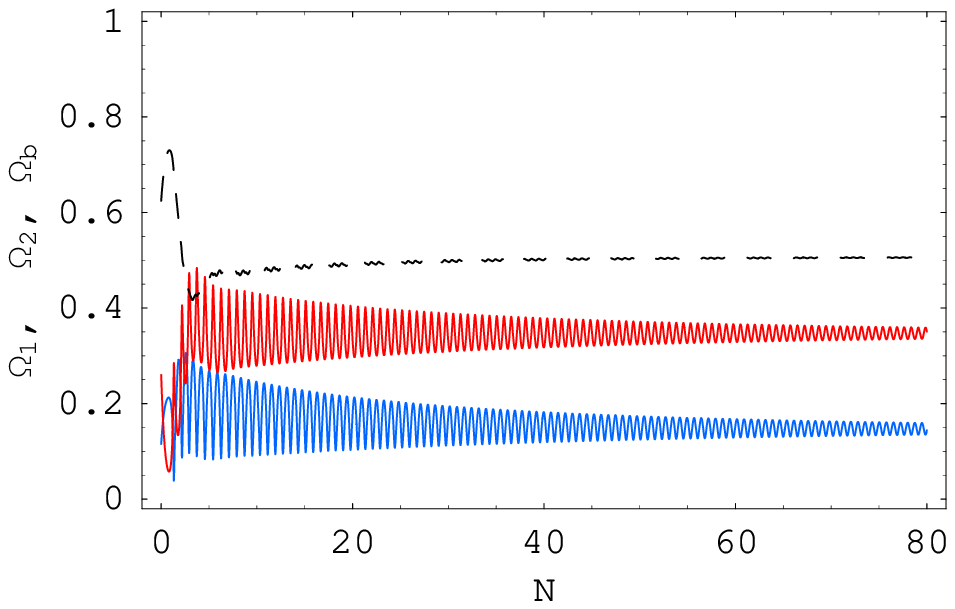}
\includegraphics[width=6.5cm]{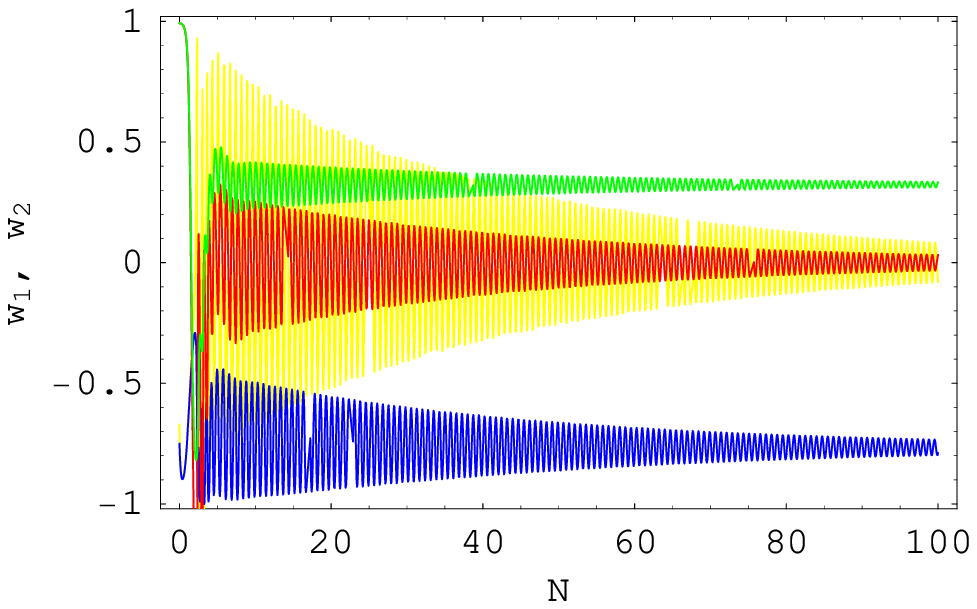}
\includegraphics[width=6.5cm]{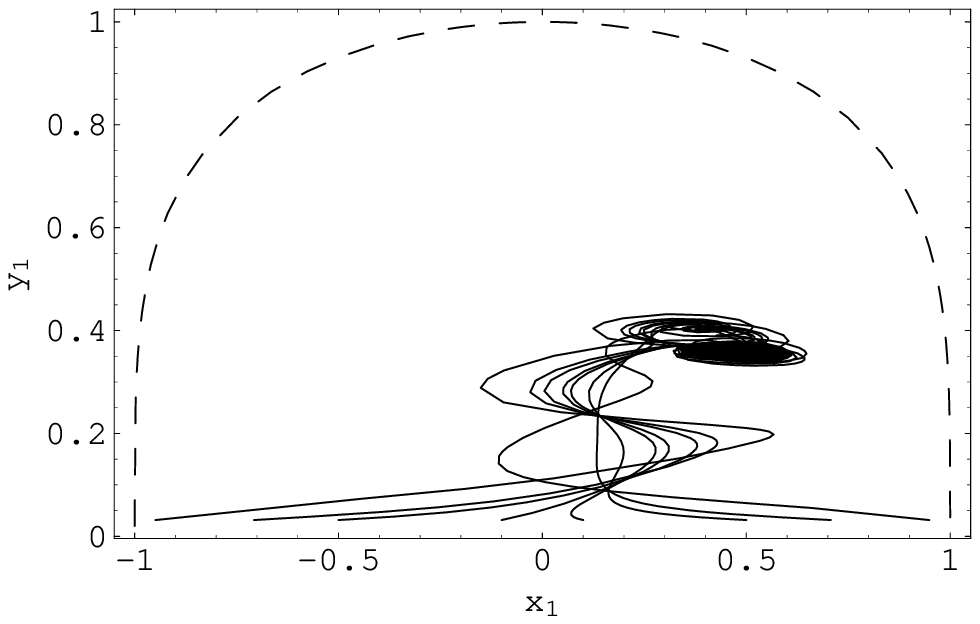}
\includegraphics[width=6.5cm]{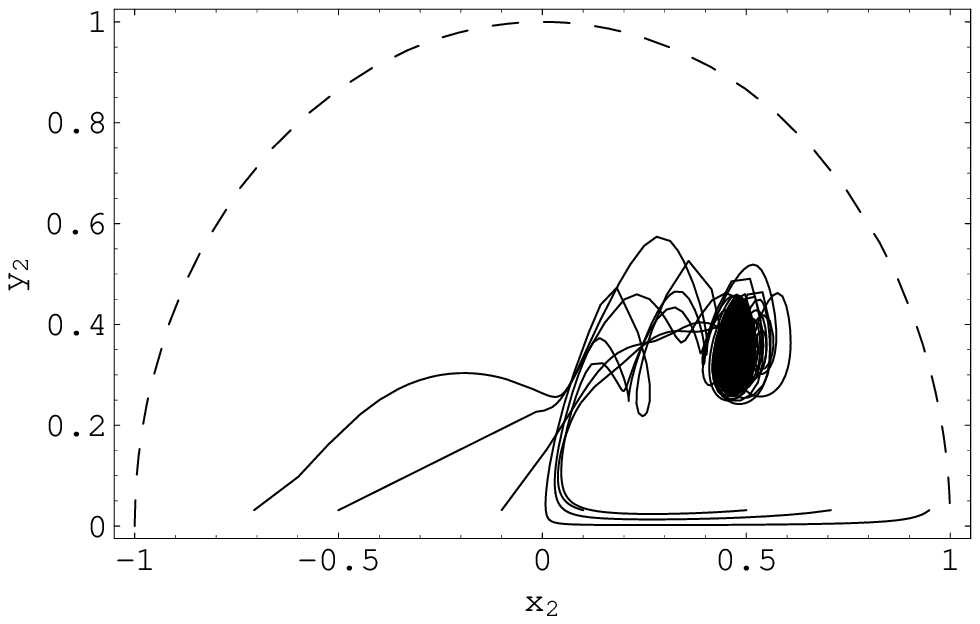}
\end{center}
\caption{\small{We show  for
$\lu=10,\ld=-10,\lt=5$ and $\gb=1+\wb=1$ the evolution of  $ \Om_1=\Omt,
 \Om_2=\Omp, \Ob$ (blue (solid), red (dotted)  and black (dashed),
respectively). We also show the equation of state parameters $ w_1=w_T,\wte$
(blue,  yellow, respectively)  and   $ w_2=\wp,  \wpe$  (green, red, respectively) as a function of $N=Log[a]$.
  With these choice of
$\lm's$ the attractor solution has   $(x_1,y_1)=(0.48,0.36)$ and
$(x_2,y_2)=(0.48,0.34)$, $\Om_1=0.14,\Om_2=0.35,\Ob=0.51$
and $\weu=\wed=w_b=0$. }}
\la{fig7}
\end{figure}

\begin{figure}[htp!]
\begin{center}
\includegraphics[width=6.5cm]{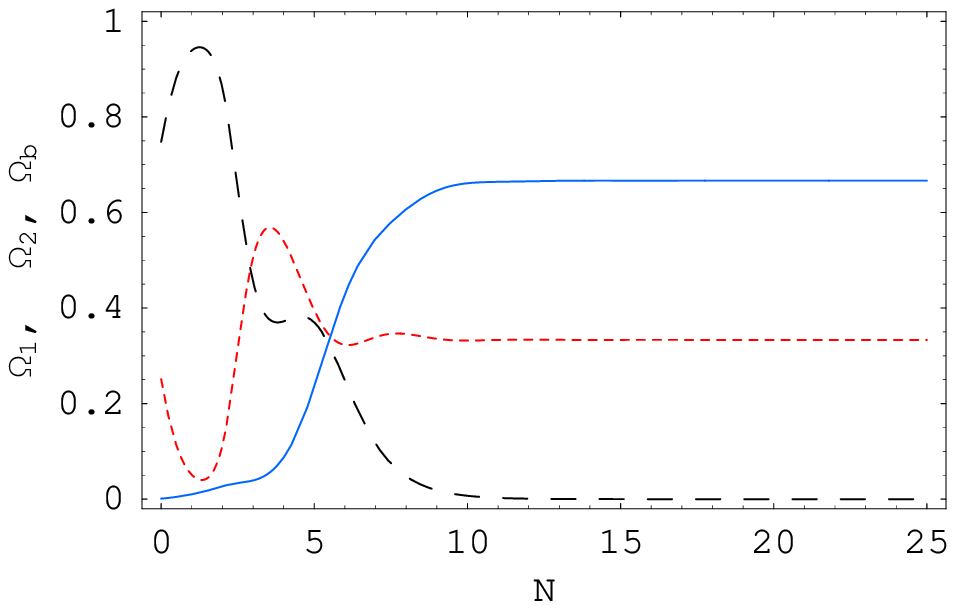}
\includegraphics[width=6.5cm]{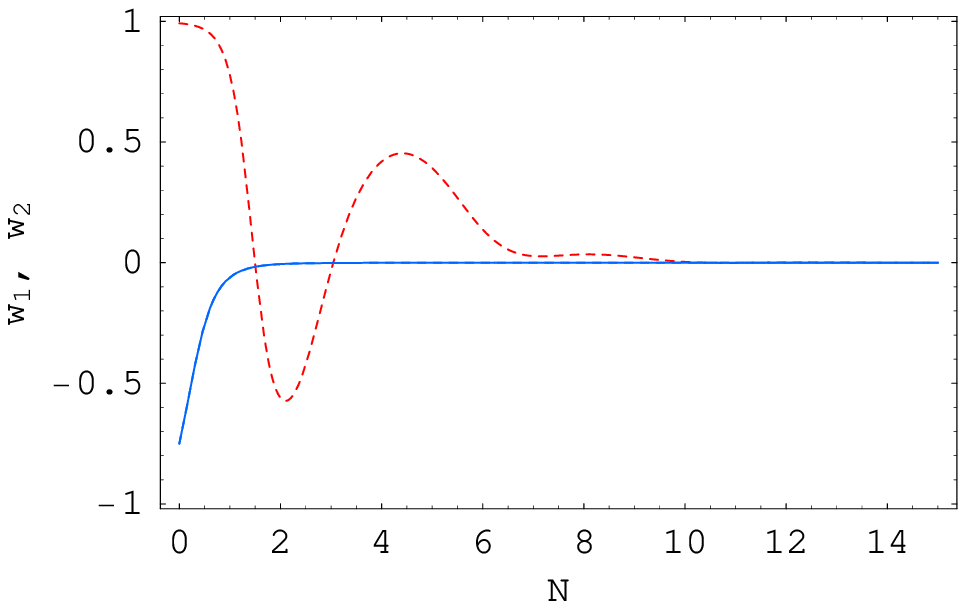}
\includegraphics[width=6.5cm]{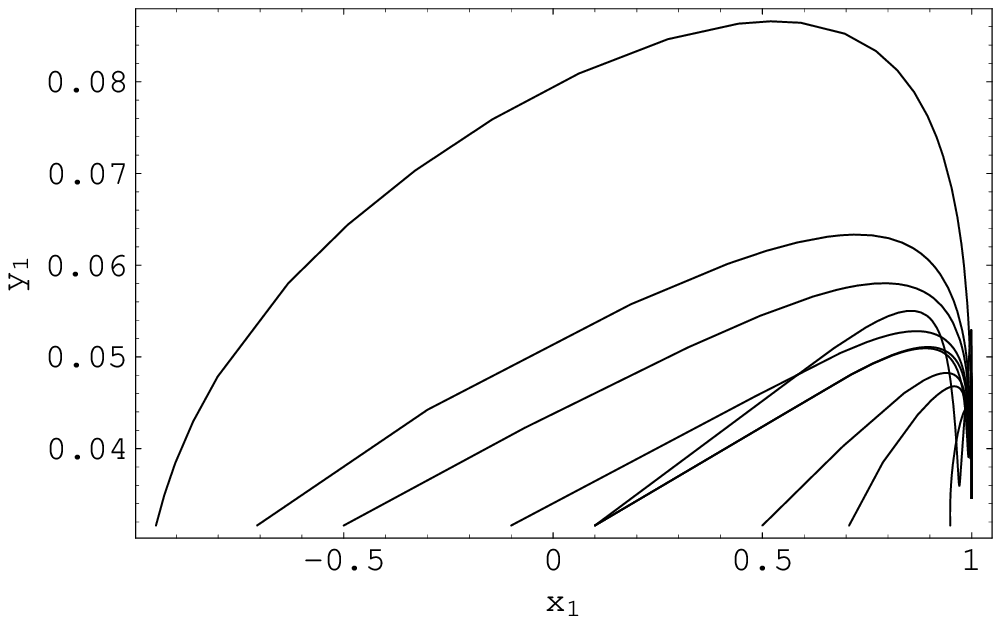}
\includegraphics[width=6.5cm]{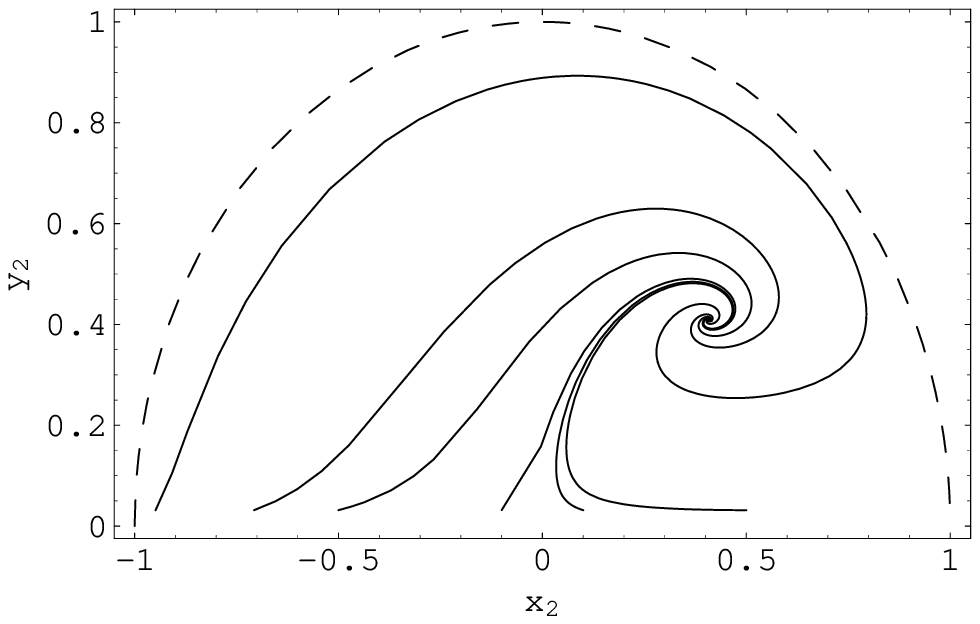}
\end{center}
\caption{\small{We show   for
$\lu=50,\ld=0,\lt=3$ and $\gb=1+\wb=4/3$ the evolution of  $ \Om_1=\Omt,
 \Om_2=\Omp, \Ob$ (blue (solid), red (dotted)  and black (dashed),
respectively). We also show   the equation of state parameters $ w_1=w_T=\wte$
(blue)  and   $ w_2=\wp=\wpe$  (red (dotted)) as a function of $N=Log[a]$.
 With these choice of
$\lm's$ the attractor solution has   $(x_1,y_1)=(0.99,0.03)$ and
$(x_2,y_2)=(0.40,0.40)$, $\Om_1=2/3,\Om_2=1/3,\Ob=0$
and $\weu=\wed=0$. }}
\la{fig5B}
\end{figure}

\begin{figure}[htp!]
\begin{center}
\includegraphics[width=6.5cm]{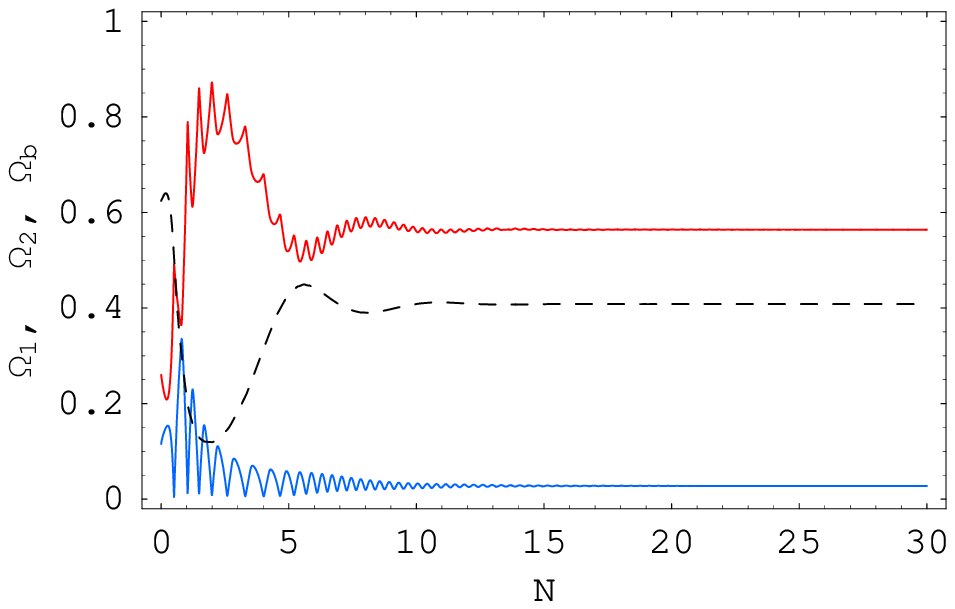}
\includegraphics[width=6.5cm]{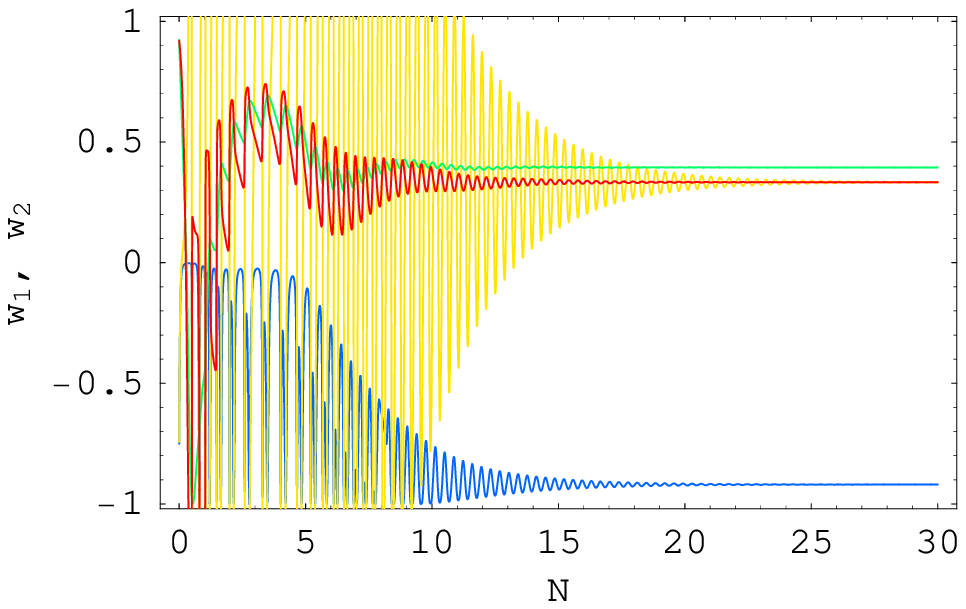}
\includegraphics[width=6.5cm]{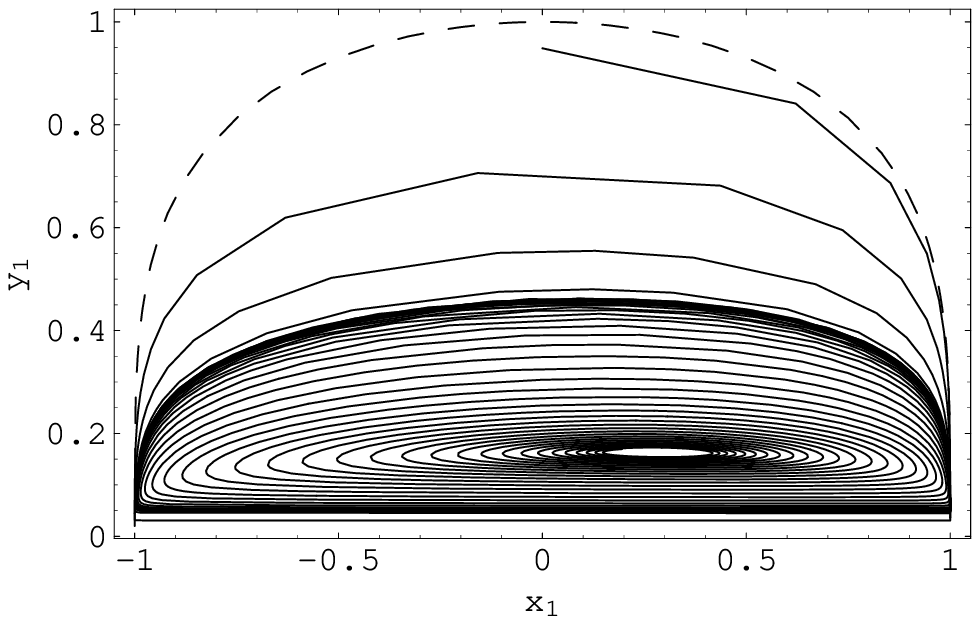}
\includegraphics[width=6.5cm]{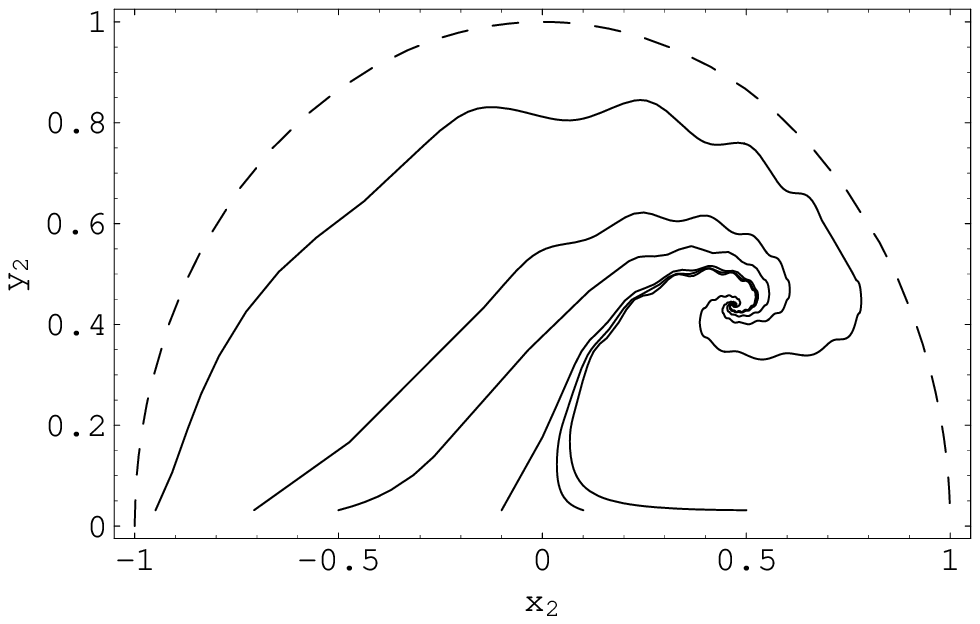}
\end{center}
\caption{\small{We show   for
$\lu=50,\ld=-3,\lt=3$ and $\gb=1+\wb=4/3$ the evolution of  $ \Om_1=\Omt,
 \Om_2=\Omp, \Ob$ (blue (solid), red (dotted)  and black (dashed),
respectively). We also show the equation of state parameters $ w_1=w_T,\wte$
(blue,  yellow, respectively)  and   $ w_2=\wp,  \wpe$  (green, red, respectively) as a function of $N=Log[a]$.
 With these choice of
$\lm's$ the attractor solution has   $(x_1,y_1)=(0.28,0.16)$ and
$(x_2,y_2)=(0.63,0.41)$, $\Om_1=0.03,\Om_2=0.56,\Ob=0.41$
and $\weu=\wed=w_b=1/3$. }}
\la{fig5}
\end{figure}

\begin{figure}[htp!]
\begin{center}
\includegraphics[width=6.5cm]{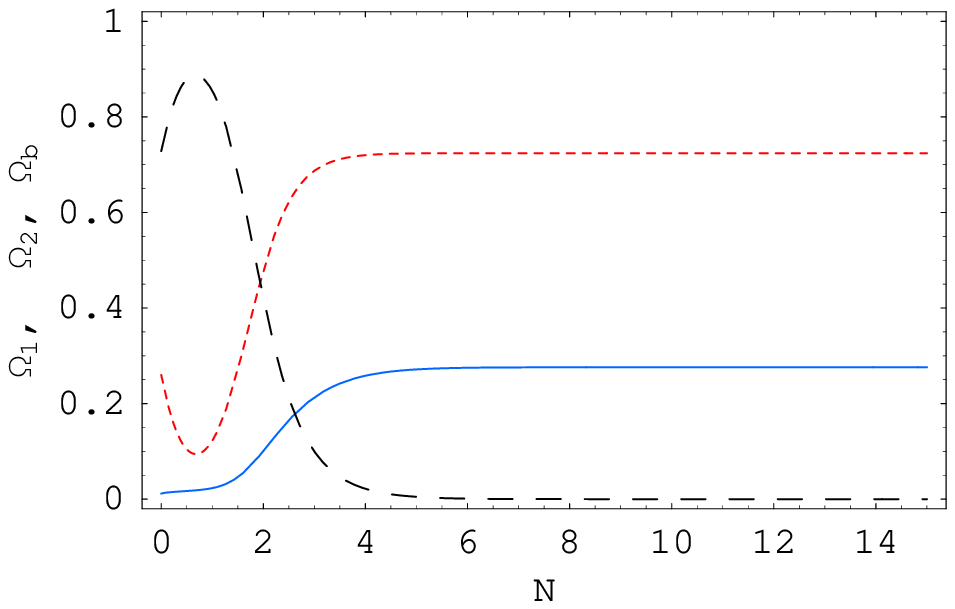}
\includegraphics[width=6.5cm]{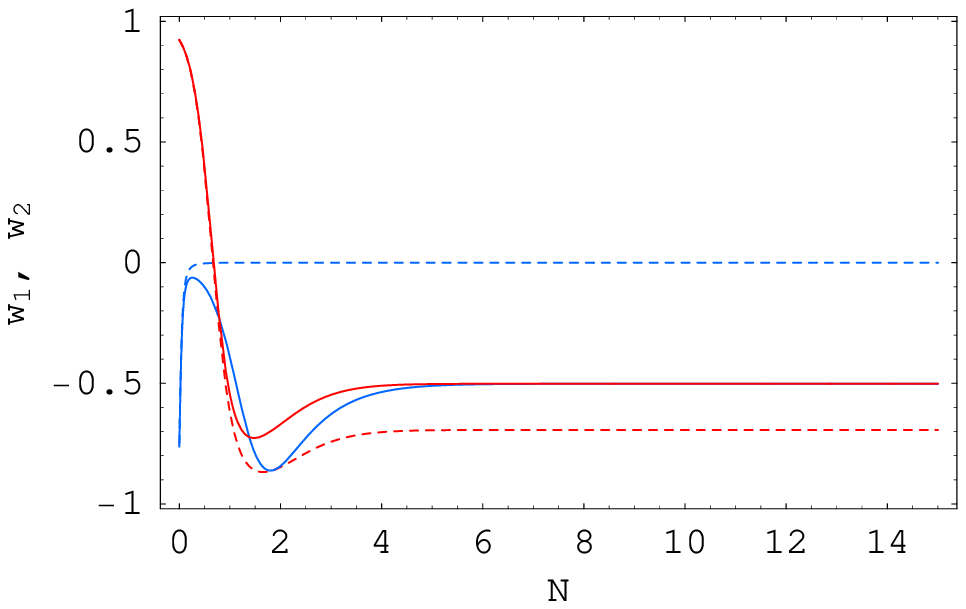}
\includegraphics[width=6.5cm]{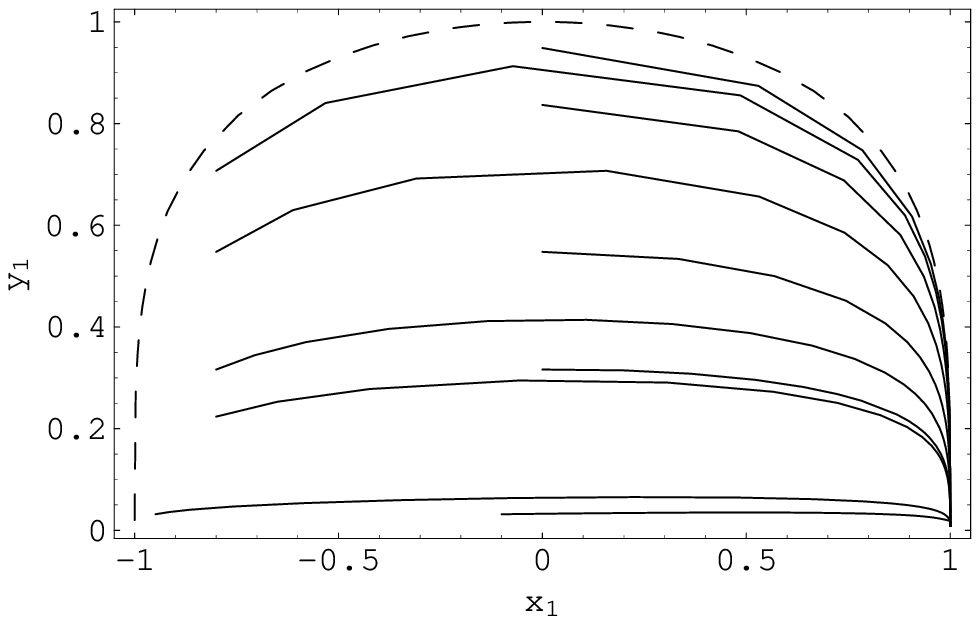}
\includegraphics[width=6.5cm]{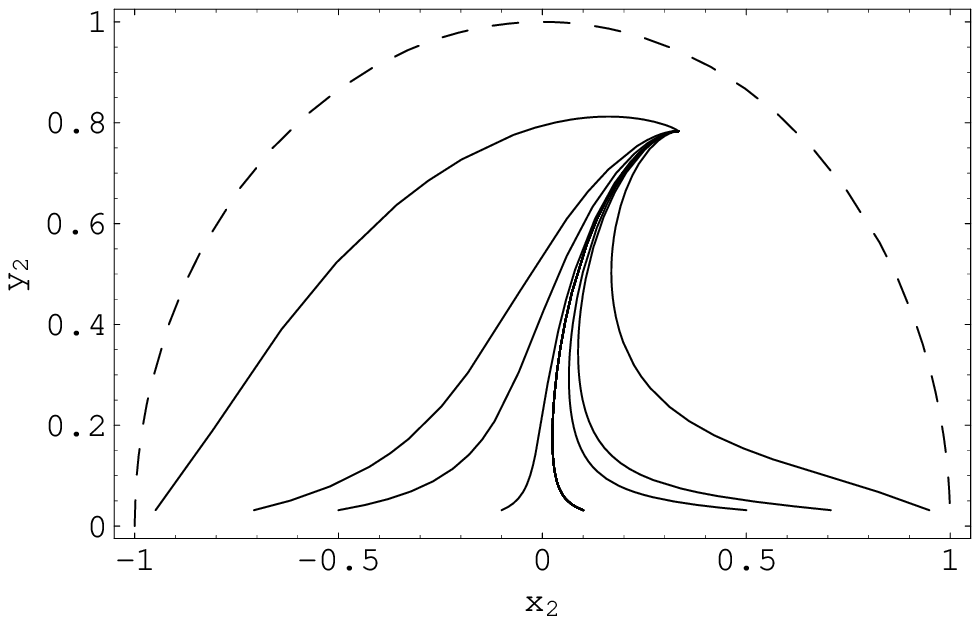}
\end{center}
\caption{\small{We show  for
$\lu=100,\ld=1/2,\lt=1$ and $\gb=1+\wb=1$ the evolution of  $ \Om_1=\Omt,
 \Om_2=\Omp, \Ob$ (blue (solid), red (dotted)  and black (dashed),
respectively). We also show the equation of state parameters $ w_1=w_T,\wte$
(blue (dotted), blue (solid), respectively)    and  $ w_2=\wp,  \wpe$   (red
(dotted)), red (solid ), respectively).
  With these choice of
$\lm's$ the attractor solution has   $(x_1,y_1)=(1,0)$ and
$(x_2,y_2)=(1/3,0.78)$, $\Om_1=0.27,\Om_2=0.73,\Ob=0$
and $\weu=\wed=w_b=-0.5$. }}
\la{fig4}
\end{figure}

\begin{figure}[htp!]
\begin{center}
\includegraphics[width=6.5cm]{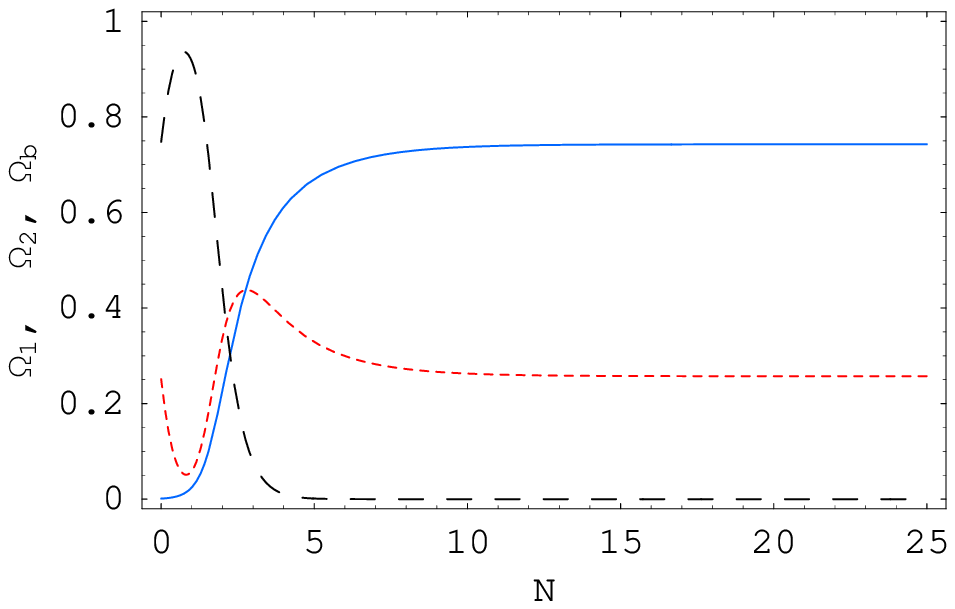}
\includegraphics[width=6.5cm]{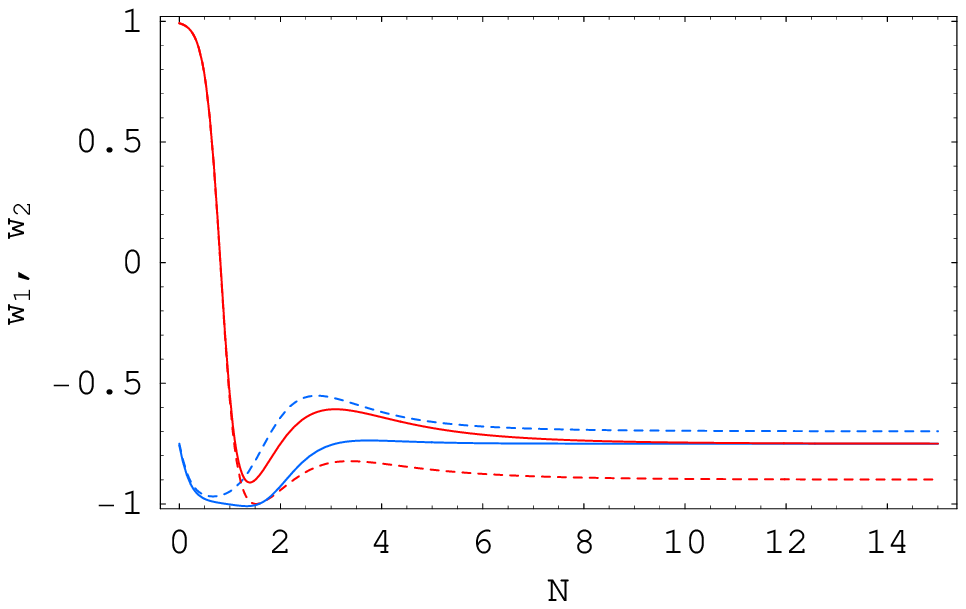}
\includegraphics[width=6.5cm]{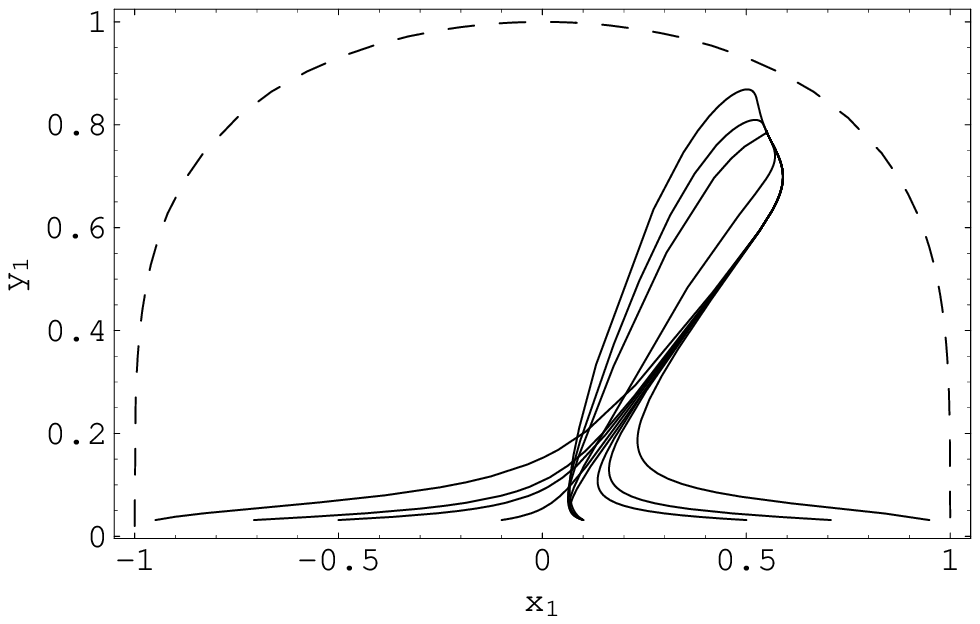}
\includegraphics[width=6.5cm]{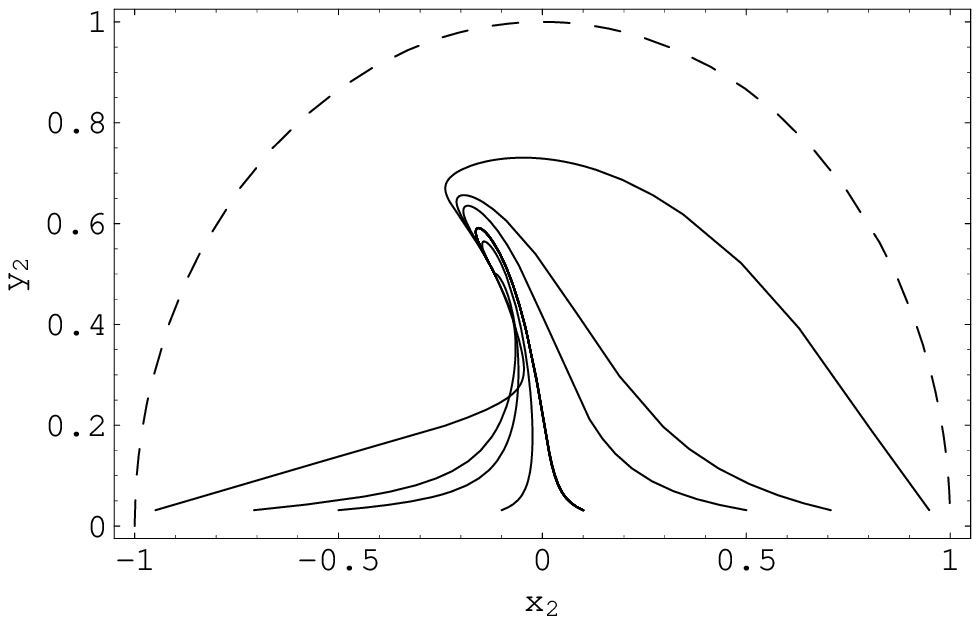}
\end{center}
\caption{\small{We show for
$\lu=1,\ld=1,\lt=-1$ and $\gb=1+\wb=1$ the evolution of  $ \Om_1=\Omt,
 \Om_2=\Omp, \Ob$ (blue (solid), red (dotted)  and black (dashed),
respectively). We also show the equation of state parameters $ w_1=w_T,\wte$
(blue,  yellow, respectively)  and   $ w_2=\wp,  \wpe$  (green, red, respectively)
 as a function of $N=Log[a]$.  With these choice of
$\lm's$ the attractor solution has   $(x_1,y_1)=(0.55,0.8)$ and
$(x_2,y_2)=(-0.11,0.49)$, $\Om_1=0.74,\Om_2=0.26,\Ob=0$
and $\weu=\wed=-0.75$. }}
\la{fig1}
\end{figure}

\section{Critical Solutions}\la{cs}

To find the critical solutions to   the dynamical equations (\ref{cosmo1})
we need to solve them
for $x_{1N}=x_{2N}=y_{1N}=y_{2N}=0$    for constant
values of $\lm_i, i=1,2,3$. However, the set of equations
are highly non linear and it is quite complicated to find
all critical points.
Instead, we will calculate the
critical points when all three fluids have the same
redshift. We will also study different limits and  we solve
eqs.(\ref{cosmo1}) numerically to show the behavior of
the scalar fields $T$ and $\phi$.

As discussed in section \ref{eff} the energy density $\r_1,\r_2,\rb$ with the smallest
effective equation of state, or alternatively with the smallest
$\gm_{i\;eff }= 1+w_{i eff},\; i=1,2,b$, will dominate the universe.
For $\gm_{i \;eff}<\gm_{k \;eff},\gm_{j\; eff}$
we have $\Om_i=1$ and $\Om_k=\Om_j=0$ at late times. If two or the three
energy densities have the same (smallest) effective equation of state then
two  $\Om$'s or all three $\Om's$ will be different than zero. Therefore,
eq.(\ref{hh}) gives asymptotically
\be\la{hh2}
-\fr{2H_N}{3H}=\le(  \Om_1\gm_1+\Om_2\gm_2 + \Ob\gb    \ri)=
\le(  \Om_1\gmeu+\Om_2\gmed + \Ob\gb    \ri)= \gme
\ee
where the last equality in eq.(\ref{hh2}) is valid only asymptotically
and $\gme=1+\we$ is the smallest of the $\gm_{eff}$'s, i.e.
$\gme=smallest(\gmeu,\gmed,\gb)$.

 If one of the scalar fields dominates over the other one
 then eqs.(\ref{cosmo1}) will be reduced to a single scalar
 field, either $T$ or $\phi$, in the presence of a barotropic
 fluid. These special case have been studied in the literature
 for tachyon domination \ci{tach.tr} and for canonical
 scalar field in \ci{mio.gen}.

So, let us  now  determine under which conditions do
both fields $T$ and $\phi$ have the same redshift. We see from
eqs.(\ref{weff2}) that both effective equations of state are equal
if
\bea\la{lde}
\fr{\ld x_1 y_2^3}{\sqrt{3}} &=& \fr{\Om_1\Om_2}{\Om_1+\Om_2}(w_1-w_2)\\
&=&\Om_1(\fr{3H_N}{2H} + \gm_1)= -\Om_2(\fr{3H_N}{2H}+\gm_2)\non
\eea
and using eq.(\ref{hh2}) and eq.(\ref{lde}) we get
\be
\fr{\Om_1}{\Om_2}=\fr{\gm_1-\gme}{\gme-\gm_2}=\fr{w_1-\we}{\we-w_2}
\ee
which coincides with eq.(\ref{OO}).
The solution of the dynamical equations given in  eqs.(\ref{cosmo1}) for finite values of $x_2,y_1,y_2$
as a function of  $x_1$ are
\bea\la{y1}
y_1&=&\fr{\sqrt{3}\gme}{\lu x_1}\\ \la{x2}
x_2&=&\sqrt{\fr{2}{3}}\fr{\lt y_2^2}{(2 -\gme)}\\ \la{y2}
y_2&=&\le(\fr{3\sqrt{3}\gme^2 (x_1^2-\gme)}{\lu^2\ld x_1^3\sqrt{1-x_1^2}}\ri)^{1/3}
=- \fr{\sqrt{3}\gme^{2/3}  }{\lu^{2/3}\ld^{1/3}  } G(x_1)
\\ \la{x1}
x_1&=& \fr{2}{\sqrt{3}\ld y_2}\le(\fr{3\gme}{2}-\fr{\lt^2 y_2^2}{2-\gme}\ri)
\eea
for $y_{1 N}=0, x_{2 N}=0,x_{1 N}=0, y_{2 N}=0$, respectively. We have defined
\be\la{g}
G(x_1)\equiv \fr{1}{x_1}\le(\fr{(\gme-x_1^2)}{\sqrt{1-x_1^2}}\ri)^{1/3}
\ee
and we obtain
\be\la{om1}
\Om_1=\fr{3\gme^2}{\lu^2x_1^2\sqrt{1-x_1^2}}
\ee
where we have used eq.(\ref{y1}) in eq.(\ref{om1}) and (\ref{y2}). The values of $y_1,x_2,y_2$ are functions
of $x_1$ and $\gme$. Solutions with $\Om_1\neq 0, \Om_2\neq 0,\Om_b\neq 0$
imply that $\gme=\gb=\gmeu=\gmed$ is a given constant (given by
the barotropic fluid). If however, $\gme=\gmeu=\gmed<\gb$
then $\Ob=0$ and $\gme$ is not a  constant given a priori but it is a constant
which is a function of $x_i,y_i$. In this case eqs.(\ref{y1})-(\ref{x1}) must be supplemented with
eq.(\ref{hh2}). Using eqs.(\ref{y1})-(\ref{y2}) we can rewrite  eq.(\ref{hh2}) as
\be\la{1}
1=\fr{\Om_1\gm_1}{\gme}+\fr{\Om_2\gm_2}{\gme}=
\fr{3\gme}{\lu^2\sqrt{1-x_1^2}}+\fr{ 3^3 4\,\lt^2 \gme^{5/3}}{(2-\gm)^2} G(x_1)^4
\ee
which is only valid if $\Ob=0$.

Let us take $\gme=\gb$ with  a  barotropic
fluid with $\gb\geq 1$, i.e. it includes
 matter $\gb=1$ and  radiation $\gb=4/3$, and in this case we get
the  following constraints on
$\lu,\ld,\lt$, from requiring $|x_i|<1$ and $ 0<y_1<1$:
\bea\la{lu}
\lu^2 &>& \fr{3\gme^2}{x_1^2\sqrt{1-x_1^2}}> \fr{9\sqrt{3}\gme^2}{2}\\
\la{ld}
|\lu^2\ld|&>&  | \fr{3\sqrt{3}\gme^2 (\gme-x_1^2)}{ x_1^3\sqrt{1-x_1^2}}|>3\sqrt{3}\gme^2 G^3_m \\
 \la{luld}
\ld x_1 &<& 0, \hspace{.3cm} \lu x_1>0,  \hspace{.3cm} \lt x_2 >0,  \hspace{.3cm} \lu\ld < 0\\
\la{lt}
\fr{\lu^{4/3}\ld^{2/3} (2-\gme)\alpha}{2\gme^{1/3}G_m^2}   &>& \lt^2 > \fr{3 \gme(2-\gme)\alpha}{2}
\eea
where $G_m$ in eqs.(\ref{ld}) and (\ref{lt}) is the minimum value
of $G(x_1)$. The constraint (\ref{lu}) comes from requiring $\Om_1<1$ and
$(x_1^2\sqrt{1-x_1^2})^{-1}$ has
a minimum value $ (x_1^2\sqrt{1-x_1^2})^{-1}= 3\sqrt{3}/2$ at $x_1^2=2/3$.
Eq.(\ref{ld}) comes  from eq.(\ref{y2})  and $0<y_2<1$.
Eq.(\ref{luld}) arises
from requiring $0<y_1$ and $ 0<y_2$  and we see from eq.(\ref{y2}) that
 $G\ld $ must be negative, which for $\gme\geq1 $ implies that $\ld x_1<0$.
Finally eq.(\ref{lt}) comes from eq.(\ref{y22}) and $ 0<y_2<1$.

For $\gb=\gme=1$ the function $G$ is  simply $G=(1-x_1^2)^{1/6}/x_1$ and $G$ takes the value
$|G(0)|=\infty$ and $G(\pm 1)=G_m=0$
so that $0\leq |G(x)| \leq \infty$. Therefore, the lower and   upper constraints in eq.(\ref{ld})
and  (\ref{lt}), respectively, do not impose any limitation to $\lm_i$.
On the other hand if $\gb=\gme>1$ (as for radiation $\gb=4/3$) then
$|G(\pm 1)|=|G(0)|=\infty$ and $G(x)$ has a minimum absolute value $G_m\equiv |G(x=x_m)|$ at
$x_m^2=[1+4\gme-\sqrt{1+16\gme(\gme-1)}]/2$, e.g.
for $\gme=4/3$ we have $x_m=0.93$ and $G_m=1.16$. The fact that $G$ has a minimum value
for $\gme=\gb>1$ implies a constraint on the value of $\lu,\ld$ namely
$\lu^2\ld> 3\sqrt{3}\gme^2 G^3_m$ as seen from eq.(\ref{ld})
and from eq.(\ref{lt}) an upper   value for $\lt$ .

To finally solve eqs.(\ref{y1})-(\ref{x1})   we substitute $y_2$ from eq.(\ref{y2}) into
eq.(\ref{x1}) and we get a nonlinear equation  for $x_1$ only
\be\la{y22}
y_2^2=  \fr{3\gme^{4/3}}{\lu^{4/3}\ld^{4/3}} G(x_1)^2
= \fr{3(2-\gme)}{2\lt^2}\le(\gme - \fr{\ld x_1 y_2}{\sqrt{3}} \ri)\equiv
\fr{3(2-\gme)\gme \alpha}{2\lt^2}
\ee
with
\be\la{al}
\alpha\equiv 1+\le(\fr{\ld}{\lu}\ri)^{2/3}\gme^{-1/3}x_1 G(x_1)
\ee
and $\alpha$ depends on  $\ld/\lu$ and $x_1$ with  $\alpha>1$ since the second term
in eq.(\ref{al}) is positive for $\gme \geq 1$. Eq.(\ref{y22}) is however valid for
any value of $\gme $ including $\gme<1$. In this later case the second
term in eq.(\ref{al}) is no longer positive definite ($x_1G $ is negative for $x_1^2> \gme$)
and the fact
that $\alpha$  must be non negative since $y_2^2>0$, gives the constraint
$\ld^{2/3}\lu^{-2/3}\gme^{-1/3}x_1 G(x_1)>-1$.

The solution to eq.(\ref{y22}) is non analytical
and must be solved numerically for $x_1$. In terms of $G$ eq.(\ref{y22}) gives
\be\la{g2}
G(x_1)^2=\fr{\lu^{4/3}\ld^{2/3}}{ \lt^2 }\fr{\alpha(2-\gme)  }{2\gme^{1/3}} \geq  G_m^2.
\ee
For $\gb=\gme=1$ eq.(\ref{g2}) does not impose a constraint on the choice of $\lu,\ld,\lt$ since
$G$ can take values form zero to infinity and $G_m=0$. However,
for $\gme >1 $ the function $G$ has a minimum value given by $G_m$, therefore the choice
of $\lu,\ld,\lt$ must be such as to give a value of $G$ larger than $G_m$
and for $\gb=4/3$, as mentioned before,  we have $G_m=1.16$.

Regarding the quantity $\alpha$ given in eq.(\ref{al})
which term dominates  (first or second)  depends on the value of $\ld/\lu$ and $x_1$
through the function  $F(x_1)=x_1 G(x_1)=((\gme-x_1^2)/\sqrt{1-x_1^2}\;)^{1/3}$.
The function $F$ has a minimum value
at $x_1=2-\gme$ giving a value $F|_m=2^{1/3}(\gme-1)^{1/6}$. For $\gme=1$ one has
$0\leq F=\sqrt{1-x_1^2}  \leq 1$
while  for $\gme >1$ one finds $F|_m\leq F\leq F(x=1)=\infty$. However, since
$F(0)=F(\widetilde{x}_1)=\gme^{1/3}$ for $\widetilde{x}_1=\sqrt{ \gme(2-\gme)}$
 the value of $F$ in the  case $\gme>1$ is constraint
between $F_m\leq F< F(0)=F(\widetilde{x}_1)=\gme^{1/3}$ for $0<|x_1|< \sqrt{ \gme(2-\gme)}$,
e.g. for $\gme=4/3$
we have $F_m=1.05\leq F\leq  F(0)=(4/3)^{1/3}=1.1$
for  $0<|x_1|<2\sqrt{2}/3=.94$ which covers  most of the range $|x_1|=(0,1)$.  Therefore,
 $\gme^{-1/3}F$  is
in general of order unity giving $\alpha$ in eq.(\ref{al}) as
\be
\alpha\simeq 1+(\ld/\lu)^{2/3}.
\ee
If we have   $\gme=\gmeu=\gmed<\gb$
then $\Ob=0$ and $\gme$ is not a given constant but is a function
of $x_i,y_i$ and may be smaller than 1. In this case the solution to eqs.(\ref{y1})-(\ref{x1}) is given
by solving eqs.(\ref{1}) and (\ref{g2}) numerically. If $\gme<1$ then the minimum
of $|G|$ and $|F|$ are  $G_m=F_m=0$ with  $0\leq G\leq \infty$ and $0\leq F\leq \infty$
and there is no upper constraint on $\lt$ from eq.(\ref{lt}) and no lower constrain on
$|\lu^1\ld|$ in eq.(\ref{ld}).
However, since  $F(x_1)=x_1 G(x_1)=((\gme-x_1^2)/\sqrt{1-x_1^2}\;)^{1/3}$  can now  be
negative  for $x_1^2> \gme$ the condition $y_2>0$ (i.e. $\alpha>0$) implies from
eq.(\ref{al})   the constraint
\be
F(x_1)^3= \fr{\gme-x_1^2}{\sqrt{1-x_1^2}}  > -\fr{\lu^{2}\gme}{\ld^{2}}.
\ee

\section{Asymptotic Behavior}\la{lim}

Special cases can be studied by taking different limits of the parameters $\lm_i$.
A constant $\lu$ is given by a potential $V=V_o/T^2$ and we
have  $\lu=2/\sqrt{V_o}$. A constant
$\lt$   can be obtained for example  if we consider  a
factorizeable  interaction potential $B(\phi,\vp)=h(T)K(\phi)$.
In this case $\lt=-K_\phi/K$  and
taking $K(\phi)$ as an exponential potential we have  $\lt=-\beta$
for $K=K_o\,e^{\beta\phi}$. However, a factorizeable potential $B$
gives $\ld = -  h_T/ (h^{3/2}K^{1/2})$ which is now a function of
both fields $T$ and $\phi$. So, a constant $\ld$ can
only be obtained as a combination of the evolution of
$T,\phi$ and cannot be trivially  anticipated just
by looking at the functional form of $B$.

An interesting limit is when $x_1^2=\dot T^2=1$. In this case
$y_1=0$ since $\Om_1=y_1^2/\sqrt{1-x_1^2}$ must be smaller than
one. The value of $\Om_1$ depends on the value
of the parameters $\lm_i$. Solving eqs.(\ref{cosmo1})
we find two solutions.  One solution has $x_1^2=1,y_1=0$ and
\bea\la{x2y2}
x_2&=& \sqrt{\fr{3}{2}}\fr{1}{4\lt^2}\le((2-\gme)\ld^2+
4\gme \lt^2-\ld sgn[x_1]\sqrt{8(2-\gme) \gme\lt^2+(2-\gme)^2\ld^2}\ri)\non\\
y_2&=&  \fr{\sqrt{3}}{4\lt^2}\le(-(2-\gme)\ld sgn[x_1]  +\sqrt{8(2-\gme)\gme\lt^2+(2-\gme)^2\ld^2}\ri)
\eea
where $sgn[x_1]=\pm 1$ and  eqs.(\ref{x2y2})
constrains the values of $\ld,\lt$ since $|x_2|<1, 0<y_2<1$. We
see that this solution requires $|\ld/\lt|<1$.
The second solution  has $x_1^2=1,y_1=0$  and $(x_2,y_2)=(0,0)$, i.e. $\Om_2=0$,
which reduce to the case without canonical scalar field
discussed  below.

In the limit $\ld=0$ the eqs.(\ref{cosmo1}) reduce to two uncoupled scalar
fields  with  a vanishing interaction term $\delta$   and
$\gmeu=\gm_1,\, \gmed=\gm_2$.  The solution can be obtain from the combination
of the work  given in \ci{tach.tr} and \ci{mio.gen}, where a tachyon
field and a scalar field in the presence of a barotropic fluid, respectively,
are studied.  Which term dominates at late times depends on  which equation of state
is smaller and this depends on the
values of $\lu,\lt$. The equation of state for  the field
$T$ is given by $\gmeu=\gm_1= (\sqrt{\lu^4+36}-\lu^2)\lu^2/18$ if $\gb>\gm_1$
and $\gmeu=\gm_1=\gb$ otherwise \ci{tach.tr}
while for the field $\phi $ we have $\gmed=\gm_2=\lt^2/3$ for $\lt^2<3\gb$
and $\gmed=\gm_2=\gb$ if $\lt^2>3\gb$   \ci{mio.gen}.
Therefore if $\lt^2<  (\sqrt{\lu^4+36}-\lu^2)\lu^2/6$ then
we have $\gm_2<\gm_1$ and the energy density $\rp$  will
dominate over $\rt$ at late times. For $\lt^2>
 (\sqrt{\lu^4+36}-\lu^2)\lu^2/6$ we have $\gm_2>\gm_1$ and the energy density $\rt$  will
dominate over $\rp$.

In the limit $\lt=0$ we have  from eqs.(\ref{cosmo1})  $x_{2 N}/x_2<0$
which implies a decreasing $x_2$ with an asymptotic limit $x_2=0$.
In this case
eqs.(\ref{cosmo1}) or equivalently   eq.(\ref{y22}) the
solution requires $\alpha=0$, i.e.    $F(x_1)=x_1 G(x_1)=-(\lu/\ld)^{2/3}\gme^{1/3}$.
Once $x_1$ has been obtained from $\alpha=0$ we can use eqs.(\ref{y1}) and (\ref{y2}) to
obtain the values of $y_1$ and $y_2$.

Finally, if we take $\lu=0$ then from $y_{1 N}=0$ in eqs.(\ref{cosmo1}) we
have $y_1 H_N/H=0$. If $H_N/H=0$ then
 $x_2=x_1=\Ob=0$ and from $x_{2 N}=0$ we conclude
 that $y_2=0$ and   the universe is dominate by a
constant tachyonic potential $\Om_1= y_1^2 =1$.  If however,
   $y_1=0$ then from $x_{1 N}=0$
 we need $x_1^2=1$ or $y_2=0$. For $x_1^2\neq 1$
the solution has $y_1=y_2=x_2=0$ with $\Om_1=\Om_2=0$
 and $\Omega_b=1$. If $x_1^2=1$ the solution
 is $y_1=0$ and $x_2,y_2$ are given in eqs.(\ref{x2y2}).

\subsection{Particle Physics Model}\la{ppm}

Let us now take a specific choice of  potentials motivated by particle
physics. We consider a
factorizeable  interaction potential
\be\la{B}
 B(\phi,\vp)=h(T)K(\phi).
 \ee
In this case the $\lm_i$ parameters
become only functions of a single field
\be\la{lm2}
\lu = - \fr{V_T }{ V^{3/2}},\;\;\;\ld
= - \fr{h_T }{ h^{3/2}K^{1/2}},\;\;\;\lt = - \fr{K_{\phi} }{ K}.
\ee
We see from eq.(\ref{lm2}) that $\lu$ is only
a function of $T$,  $\lt$ is only a function of
$\phi$ while $\ld$ depends on $T$ and $\phi$.
Let us take  from eq.(\ref{vv}) $dV_{eff}/dT=
 (V_T +B_T \sdt\;)V/\rt^2=0$. This equation has
 two solutions  given in eqs.(\ref{s1}) and (\ref{sdt}).
If $\dot T\neq 1$ then using the anstaz in eq.(\ref{B})
we have from eq.(\ref{vv})
\be\la{K}
K(\phi)=-\fr{V_T}{h_T\sdt}=-\fr{\rt V_T}{h_T V}
\ee
where we have used $\rt=V/\sdt$ and substituting $K$ in eq.(\ref{lm2})
we get
\be
\ld=\le(\fr{h_T}{h}\ri)^{3/2}\le(-\fr{V}{V_T}\ri)^{1/2}\fr{1}{\rt^{1/2}}.
\ee
The solution in eq.(\ref{sdt}) has $\dot T\simeq 0$ and $ \rt\simeq V$.
As an example let us take the potential for brane-tachyon field
as $V=V_o\,e^{-T^2/2}$ and the interaction term
as $B=B_o e^{\beta T}\phi^n$, i.e. $h=e^{\beta T}$
and $K=B_o \phi^n$,  with $V_o>0, B_o>0$. In this
 case we have from eq.(\ref{K}) $K=\rt T e^{-\beta T}/\beta$
 \be
 \lu=T e^{T^2/4},\;\;\; \ld=-\fr{\beta^{3/2}}{ (T\rt)^{1/2}},\;\;\;
 \lt=-\fr{n}{\phi}=\pm nK^{-1/n}=\pm \fr{n\beta^{1/n} e^{\beta T/n}}{(T\rt)^{1/n}}
 \ee
and in the limit $T\rightarrow\infty$ we have
\be\la{lm6}
 \lu \rightarrow\infty ,\;\;\;\;\;  \ld \rightarrow -\infty,\;\;\;\;\;
 \lt  \rightarrow \pm \infty
 \ee
where we have taken $\beta>0$ as required by eq.(\ref{vv}) since
$V_T$ and $B_T$  must have opposite signs. The relative
growth is given by
\be\la{lm7}
 \fr{\lu}{\ld}\simeq T^{3/2} , \;\;\;\;\;
 \fr{\lt}{\lu}\simeq T^{-(n+1)/n}e^{\beta T +T^2(2-n)/4n}, \;\;\;\;\;
\fr{\lt}{\ld}\simeq T^{(n-2)/n}e^{\beta T +T^2(2-n)/4n}.
 \ee
Form eq.(\ref{lm7}) we have $\lu\gg |\ld|$ and for $n=2$
we get $|\lt|\gg\lu\gg|\ld|$ while for $n>2$ we find
$\lu\gg|\ld|\gg|\lt|$. On the other hand if $\dot T^2=1$ then we have $V=0$ and
the solution has either $\Om_2=0$ or it is given by eq.(\ref{x2y2}).
Solving numerically
eqs.(\ref{dp}) and  (\ref{dvp})
with $n=4, \beta=1, \gb=1$ we find  $\dot T^2 \rightarrow 1$
with $|\lu|\gg|\ld|\gg|\lt|\gg 1$ and $\Om_2=\Ob=0$.

\subsection{Examples}\la{ex}

We now present four  different attractor solution depending on the values of $\lu,\ld,\lt$.
We show in figures \ref{fig7}-\ref{fig1}  the evolution of $\Omp\equiv \Om_1, \Omvp\equiv\Om_2,\Ob$
 and  $ w_1\equiv\wp,\wpe,  w_2\equiv\wvp ,  \wvpe$ for the different choices of $\lm's$.  We also show
 the phase space of  $(x_1,y_1)$ and $(x_2,y_2)$ for each case. Since the phase space depends
 on four variables, namely  $(x_1,y_1,x_2,y_2)$ , it is no surprising that the curves in
 the two dimensional space $(x_1,y_1)$ and $(x_2,y_2)$ may cross.

In figs.\ref{fig7}
we have $\lu=10,\ld=-10,\lt=-5$ and $\gb=1+\wb=1$. The conditions in
eqs.(\ref{lu})-(\ref{lt}) are therefore   satisfied and we
expect to have $\Om_1,\Om_2,\Om_b$ different than zero.
In fact we get  $\Om_1=0.14,\Om_2=0.35,\Ob=0.51$ and
$(x_1,y_1)=(0.48,0.36)$ and
$(x_2,y_2)=(0.48,0.34)$. The effective equations of state are
$\weu=\wed=w_b=0$ giving a decelerating universe.

In figs.\ref{fig5B} we show a model with vanishing interaction term $\delta = 0$,
i.e. $\ld=0$,  and with
 $\lu=50,\lt=3$ and $\gb=1+\wb=4/3$.
The attractor solution has   $\Om_1=2/3,\Om_2=1/3,\Ob=0$ and $(x_1,y_1)=(0.99,0.03)$
$(x_2,y_2)=(0.40,0.40)$. The  equation of state $w_i$ is equal to
the  effective equation of state $w_{i eff}=w_i=0$ and $T$
redshifts as matter at late times.

In figs.\ref{fig5} we show the same  model as in figs.\ref{fig5B} but with
an interaction term $\delta\neq 0$. We take $\ld=-3$ and
 $\lu=50,\lt=3$ and $\gb=1+\wb=4/3$.
The attractor solution has
$\Om_1=0.03,\Om_2=0.56,\Ob=0.41$ and $(x_1,y_1)=(0.28,0.16)$
$(x_2,y_2)=(0.63,0.41)$.
The  effective equation of state is given by
$\weu=\wed=w_b=1/3$ which shows that the tachyon field and $\phi$
redshift as radiation.  We see
that the relevance of the interaction term in figs.\ref{fig5}
which makes $w_{1 eff}$  go from $w_{1 eff}=0 $ (no interaction)
 to $w_{1 eff}=1/3$ (with interaction). This shows that an interaction term can solve
the  problem of the tachyon field redshifting as matter
and dominating the universe   well before  radiation-matter equality.

In figs.\ref{fig4}
we have $\lu=100,\ld=1/2,\lt=1$ and $\gb=1+\wb=1$. With these choice of
$\lm's$ the attractor solution has   $(x_1,y_1)=(1,0)$ and
$(x_2,y_2)=(1/3,0.78)$, $\Om_1=0.27,\Om_2=0.73,\Ob=0$
and $\weu=\wed=w_b=-0.5$. Notice that even tough $y_1 \rightarrow 0$
we have $x_1\rightarrow 1$ and a finite $\Om_1$ which dominates
the universe with $\weu=\wed=-0.5$ giving a positive acceleration.
Notice that $w_1\rightarrow 0$ as for a tachyon with no interaction,
however the interaction term gives a negative effective equation of state.

In figs.\ref{fig1}
we have $\lu=1,\ld=1,\lt=-1$ and $\gb=1+\wb=1$. The conditions in
eqs.(\ref{lu})-(\ref{lt}) are therefore not satisfied and we
do not expect to have $\Om_1,\Om_2,\Om_b$ different than zero.
In fact we get $\Om_1=0.74,\Om_2=0.26,\Om_b=0$, i.e. the $T$ field
dominates at late times,  and $(x_1,y_1)=(0.55,0.8)$
$(x_2,y_2)=(-0.11,0.49)$.
The effective equations of state are
$\weu=\wed=-0.75$ giving an accelerating universe.

\section{Summary and Conclusions}\la{con}

We have studied the dynamical system of two
scalar fields, a tachyon and a canonically normalized
field,  with arbitrary potentials in the
presence of a barotropic fluid in a FRW metric. We have shown
that all  model dependence is given in terms
of three parameters, namely $
\lu = -  V_T/ V^{3/2},
\ld  = -  B_T / B^{3/2}$ and
$\lt =-B_{\phi} / B$.
The solution to the dynamical equations given in (\ref{cosmo1})
are non linear  and general  analytic solution does
not exist.  If one of the scalar fields dominates over the other one
 then eqs.(\ref{cosmo1}) will be reduced to a single scalar
 field, either $T$ or $\phi$, in the presence of a barotropic
 fluid. These special cases have been studied in the literature,
 for a tachyon domination in \ci{tach.tr} and for a canonical
 scalar field in \ci{mio.gen}.
We have determined the critical points
in the case where the two scalar fields $T$ and $\phi$
have the same redshift. The solution is a function
of the parameters $\lm_i$ and we have calculated the
restrictions on these parameters. We have solved numerically
for different interesting cases and we have shown
that either scalar field or the barotropic fluid
can dominate at late times. The interaction term between the tachyon and
the scalar field  changes the effective equation
of state for $T$ and   it is  possible for a tachyon field
to  redshift as matter in the absence of an interaction
term $B$   and   as radiation when
$B$ is turned on. This results solves then the tachyonic matter
problem.

\section*{Acknowledgments}\non

This work was also supported in
part by CONACYT project 45178-F and DGAPA, UNAM project
IN114903-3.


\begin{thebibliography}{99}










\bibitem{DB2}
A.~Sen,
  %
  JHEP {\bf 0204}, 048 (2002);
  A.~Sen,
  %
  JHEP {\bf 0207}, 065 (2002);
  A.~Sen,
  %
  Mod.\ Phys.\ Lett.\ A {\bf 17}, 1797 (2002);
  A.~Sen,
  Phys.\ Scripta {\bf T117}, 70 (2005);
  A.~Sen,
  JHEP {\bf 9910}, 008 (1999);
  M.~R.~Garousi,
  %
  Nucl.\ Phys.\ B {\bf 584}, 284 (2000);
  M.~R.~Garousi,
  %
  Nucl.\ Phys.\ B {\bf 647}, 117 (2002);
  M.~R.~Garousi,
  %
  JHEP {\bf 0305}, 058 (2003);
  E.~A.~Bergshoeff, M.~de Roo, T.~C.~de Wit, E.~Eyras and S.~Panda,
  %
  JHEP {\bf 0005}, 009 (2000);
  J.~Kluson,
  %
  Phys.\ Rev.\ D {\bf 62}, 126003 (2000);
  D.~Kutasov and V.~Niarchos,
  %
  Nucl.\ Phys.\ B {\bf 666}, 56 (2003);


\bibitem{slowroll}
  G.~W.~Gibbons,
  %
  Class.\ Quant.\ Grav.\  {\bf 20}, S321 (2003);
  G.~W.~Gibbons,
  %
  Phys.\ Lett.\ B {\bf 537}, 1 (2002);
  M.~Fairbairn and M.~H.~G.~Tytgat,
  %
  Phys.\ Lett.\ B {\bf 546}, 1 (2002);
  A.~Feinstein,
  %
  Phys.\ Rev.\ D {\bf 66}, 063511 (2002);
  S.~Mukohyama,
  %
  Phys.\ Rev.\ D {\bf 66}, 024009 (2002)
  D.~Choudhury, D.~Ghoshal, D.~P.~Jatkar and S.~Panda,
  %
  Phys.\ Lett.\ B {\bf 544}, 231 (2002);
  G.~Shiu and I.~Wasserman,
  %
  Phys.\ Lett.\ B {\bf 541}, 6 (2002);
  L.~Kofman and A.~Linde,
  %
  JHEP {\bf 0207}, 004 (2002);
  M.~Sami,
  %
  Mod.\ Phys.\ Lett.\ A {\bf 18}, 691 (2003);
  G.~N.~Felder, L.~Kofman and A.~Starobinsky,
  %
  JHEP {\bf 0209}, 026 (2002);
  S.~Mukohyama,
  %
  Phys.\ Rev.\ D {\bf 66}, 123512 (2002);
  M.~C.~Bento, O.~Bertolami and A.~A.~Sen,
  %
  Phys.\ Rev.\ D {\bf 67}, 063511 (2003);
  J.~G.~Hao and X.~Z.~Li,
  %
  Phys.\ Rev.\ D {\bf 66}, 087301 (2002);
  C.~J.~Kim, H.~B.~Kim and Y.~B.~Kim,
  %
  Phys.\ Lett.\ B {\bf 552}, 111 (2003);
  B.~C.~Paul and M.~Sami,
  %
  Phys.\ Rev.\ D {\bf 70}, 027301 (2004);
  T.~Padmanabhan,
  %
  Phys.\ Rev.\ D {\bf 66}, 021301 (2002);
  J.~S.~Bagla, H.~K.~Jassal and T.~Padmanabhan,
  %
  Phys.\ Rev.\ D {\bf 67}, 063504 (2003);
  A.~V.~Frolov, L.~Kofman and A.~A.~Starobinsky,
  %
  Phys.\ Lett.\ B {\bf 545}, 8 (2002);


\bib{mio.gen}
A. de la Macorra, G. Piccinelli, Phys.Rev.D61:123503,2000.
\bib{copeland} T. Barreiro, Edmund J. Copeland, N.J. Nunes,
Phys.Rev.D61:127301,2000.



\bib{tach.mio}
A. de la Macorra, U. Filobello, G. German,  Phys.Lett.B635:355-363,2006.

\bibitem{tach.tr}
  E.~J.~Copeland, M.~R.~Garousi, M.~Sami and S.~Tsujikawa,
  Phys.\ Rev.\ D {\bf 71}, 043003 (2005);
J.M. Aguirregabiria, R. Lazkoz, Phys.Rev.D69:123502,2004

\bib{mio.2gen}
A. de la Macorra  astro-ph/0703702





\end{thebibliography}
\end{document}